\documentclass[10pt]{article}
\usepackage{amssymb}
\usepackage{amsmath}
\usepackage{epsfig}
\usepackage{subfigure}
\usepackage{graphics}

\usepackage{latexsym}
\usepackage{rotating}
\usepackage{titlesec}
\usepackage{authblk}
\title{Superposition of the Coupled NLS and MKdV Systems}

\author[1]{Metin G\"{u}rses \thanks{gurses@fen.bilkent.edu.tr}}
\author[2]{Asl{\i} Pekcan \thanks{Corresponding author. aslipekcan@hacettepe.edu.tr}}
\affil[1]{\small{Department of Mathematics, Faculty of Science, Bilkent University, 06800 Ankara - Turkey}}
\affil[2]{\small{Department of Mathematics, Faculty of Science, Hacettepe University, 06800 Ankara - Turkey}}

\date{\nonumber}
\begin{document}
\maketitle
\date{}
\newtheorem{thm}{Theorem}[section]
\newtheorem{Le}{Lemma}[section]
\newtheorem{defi}{Definition}[section]
\newtheorem{ex}{Example}[section]
\newtheorem{pro}{Proposition}[section]

\numberwithin{equation}{section}

\begin{center}
\textbf{Abstract}
\end{center}
\noindent {\small{Superpositions of hierarchies of integrable equations  are also integrable.  The superposed equations, such as the Hirota equations in the AKNS hierarchy, cannot be considered as new integrable equations. Furthermore if one applies the Hirota bilinear method to these equations one obtains the same $N$-soliton solutions  of the generating equation which differ  only by the dispersion relations. Similar discussions can be made for the locally and nonlocally reduced equations as well. We give, as an example, AKNS system of equations in $(1+1)$-dimensions.

}}

\section{Introduction}

In this paper we  point out two important misunderstandings  of the system of integrable equations possessing recursion operators.
The first one is about the superposition of integrable equations. In general the integrable equations admitting recursion operator  generate an infinite hierarchy of other integrable equations. Superpositions of different members of this hierarchy are also integrable. These equations should not be proposed as new integrable equations, because they are integrable by construction. Second point is that soliton solutions obtained via the Hirota bilinear method of the superposed equations have the same soliton solutions of the generating equation (generally the first member)  of the hierarchy, the only difference is their dispersion relations.

 In $(1+1)$-dimensions, the following system
\begin{eqnarray}
&&q^{i}_{t}=F^{i}(q^{j}, r^{j}, q^{j}_{x}, r^{j}_{x}, q^{j}_{xx}, r^{j}_{xx}, \cdots),~~~ i, j=1, 2, \cdots, N  \label{cdenk1}\\
 \nonumber \\
&&r^{i}_{t}=G^{i}(q^{j}, r^{j}, q^{j}_{x}, r^{j}_{x}, q^{j}_{xx}, r^{j}_{xx}, \cdots),~~~ i, j=1, 2, \cdots, N \label{cdenk2}
\end{eqnarray}
of evolution equations is called integrable, where $F^{i}$ and $G^{i} ~(i=1,2,\cdots,N)$ are functions of the dynamical variables $q^{i}(x,t)$, $r^{i}(x,t)$, and their partial derivatives with respect to $x$, if it has a Lax pair and a recursion operator $\mathcal{R}$.

 There are local and nonlocal reductions of the above system of equations (\ref{cdenk1}) and (\ref{cdenk2}). They are given as follows:

\noindent (a) Local reductions: The local reductions are given by
\begin{equation}\label{generalred1}
r^{i}(x,t)=\kappa_{1}\, q^{i}(x,t),\quad i=1, 2, \cdots, N
\end{equation}
and
\begin{equation}\label{generalred2}
r^{i}(x,t)=\kappa_{2}\, \bar{q}^{i}(x,t),\quad i=1, 2, \cdots, N,
\end{equation}
where $\kappa_{1}$ and $\kappa_{2}$ are real constants and a bar over a letter denotes complex conjugation. If the reductions are consistent the system of equations (\ref{cdenk1}) and (\ref{cdenk2}) reduce to one system for $q^{i}$'s
\begin{equation}\label{first}
q^{i}_{t}={\tilde F}^{i}(q^{j}, q^{j}_{x}, q^{j}_{xx}, \cdots), \quad i, j=1,2,\cdots, N
\end{equation}
for the reduction (\ref{generalred1}) and
\begin{equation}\label{second}
q^{i}_{t}={\tilde F}^{i}(q^{j}, \bar{q}^{j}, q^{j}_{x}, \bar{q}_x^{j}, q^{j}_{xx},\bar{q}_{xx}^j, \cdots), \quad i, j=1,2,\cdots, N
\end{equation}
for the reduction (\ref{generalred2}), where ${\tilde F}=F|_{r^{i}=\kappa_1 q^{i}}$ and ${\tilde F}=F|_{r^{i}=\kappa_2 \bar{q}^{i}}$, respectively.

\noindent {\bf (b)} Nonlocal reductions: Recently, Ablowitz and Musslimani introduced a new type of reduction \cite{AbMu1}-\cite{AbMu3} (see also \cite{GurPek1}-\cite{chen})
\begin{equation}
r^{i}(x,t)=\tau_{1} q^{i}(\varepsilon_{1} t, \varepsilon_{2} x)=\tau_1q_{\varepsilon}^i,\, \tau_1\in \mathbb{R} \label{red1}
\end{equation}
and
\begin{equation}
r^{i}(x,t)=\tau_{2} \bar{q}^{i}(\varepsilon_{1} t, \varepsilon_{2} x)=\tau_2\bar{q}_{\varepsilon}^i,\, \tau_2\in \mathbb{R}\label{red2}
\end{equation}
for $i=1,2,\cdots,N$ and $\varepsilon_{1}^2=\varepsilon_{2}^2=1$. When
$(\varepsilon_{1}, \varepsilon_{2})=\{(-1,1),(1,-1),(-1,-1)\}$ the above constraints reduce the system (\ref{cdenk1}) and (\ref{cdenk2}) to nonlocal space reflection symmetric (S-symmetric), time reflection symmetric (T-symmetric), or space-time reflection symmetric (ST-symmetric) differential equations.

Since the reductions are done consistently  the reduced systems of equations must  also be integrable. This means that the reduced systems
admit recursion operators and bi-Hamiltonian structures.  We can obtain $N$-soliton solutions of the reduced systems, for instance, by the inverse scattering method  \cite{AbMu2}, \cite{JZ2}, Darboux transformation \cite{JZ1}, and Hirota bilinear method \cite{GurPek1}-\cite{GurPek3}.

\section{Superposition of integrable equations}

We can  write (\ref{cdenk1}) and (\ref{cdenk2}) by the use of the recursion operator as
\begin{equation}\label{generalsystem}
u^{\alpha}_t=\mathcal{R}^{\alpha}\,_{\beta}\, u^{\beta}_x,\quad u^{\alpha}=\left( \begin{array}{c}
q^{i}  \\
r^{i}
 \end{array} \right)
\end{equation}
for $ \alpha, \beta=1,2,\cdots, 2N$, $i=1,2,\cdots,N$, and $\mathcal{R}$ is the recursion operator which is a $2N\times 2N$ matrix. Here repeated indices are summed over their range. In the sequel we omit the indices over the dynamical variables $u$, $q$, and $r$ for simplicity. The system in (\ref{generalsystem}) has an infinite hierarchy of equations \cite{olv}
\begin{equation}\label{generalsystem1}
u_{t_{N}}=\mathcal{R}^{N}\,\, u_x,~~~N=1,2,\cdots .
\end{equation}
Since the recursion operator is the symmetry
generating operator we can easily modify the above system of equations as the following superposition of the hierarchy of equations:
\begin{equation}\label{generalcomb}
u_{t_{n}}=a_1\mathcal{R} u_x+a_2\mathcal{R}^2 u_x+\cdots+a_n\mathcal{R}^n u_x,\quad u=\left( \begin{array}{c}
q  \\
r
 \end{array} \right),
\end{equation}
where $a_1, a_2, \cdots, a_n$ are arbitrary constants and $n=1,2, \cdots$. The system (\ref{generalcomb}) is also integrable
and its recursion operator is also $\mathcal{R}$. The system of equations given in (\ref{generalcomb}) cannot be considered as new integrable equations.

 If we apply the local and nonlocal reductions; (\ref{generalred1}), (\ref{generalred2}), (\ref{red1}), and (\ref{red2}) to the system (\ref{generalsystem}) then we obtain reduced systems possessing the recursion operators
yielding from the recursion operator $\mathcal{R}$ of the original system (\ref{generalsystem}) as $\mathcal{R}_1=\mathcal{R}_{r=kq}$, $\mathcal{R}_2=\mathcal{R}_{r=k\bar{q}}$, $\mathcal{R}_3=\mathcal{R}_{r=kq_{\varepsilon}}$, and $\mathcal{R}_4=\mathcal{R}_{r=k\bar{q}_{\varepsilon}}$, respectively. Corresponding to each reduction we have the following hierarchies of equations:
\begin{eqnarray*}
\left( \begin{array}{c}
q  \\
q
 \end{array} \right)_{t_{N}}=&&\,\,\,\,{\cal R}_{1}^{N}\,\left( \begin{array}{c}
q  \\
q
 \end{array} \right)_{x}, ~~~~~
\left( \begin{array}{c}
q  \\
\bar{q}
 \end{array} \right)_{t_{N}}=\,\,\,\,\,{\cal R}_{2}^{N}\,\left( \begin{array}{c}
q  \\
{\bar q}
 \end{array} \right)_{x}, \\
\left( \begin{array}{c}
q  \\
q_{\varepsilon}
 \end{array} \right)_{t_{N}}=&&\,\,\,{\cal R}_{3}^{N}\,\left( \begin{array}{c}
q  \\
q_{\varepsilon}
 \end{array} \right)_{x},\hspace{0.5cm}
\left( \begin{array}{c}
q  \\
\bar{q}_{\varepsilon}
 \end{array} \right)_{t_{N}}=\,\,\,{\cal R}_{4}^{N}\,\left( \begin{array}{c}
q  \\
\bar{q}_{\varepsilon}
 \end{array} \right)_{x},
 \end{eqnarray*}
 where $N=1,2, \cdots .$

 One can consider reductions of the superposed system (\ref{generalcomb}). Similarly the reduced systems have their own
hierarchy. In particular, nonlocal reduced system of equations have the following form:
\begin{equation}
\left( \begin{array}{c}
q  \\
q_{\varepsilon}
 \end{array} \right)=a_1\mathcal{R}_{3} \left( \begin{array}{c}
q  \\
q_{\varepsilon}
 \end{array} \right)_x+a_2\mathcal{R}_{3}^2 \left( \begin{array}{c}
q  \\
q_{\varepsilon}
 \end{array} \right)_x+\cdots+a_n\mathcal{R}_{3}^n \left( \begin{array}{c}
q  \\
q_{\varepsilon}
 \end{array} \right)_x
 \end{equation}
and
\begin{equation}
\left( \begin{array}{c}
q  \\
\bar{q}_{\varepsilon}
 \end{array} \right)=b_1\mathcal{R}_{4} \left( \begin{array}{c}
q  \\
\bar{q}_{\varepsilon}
 \end{array} \right)_x+b_2\mathcal{R}_{4}^2 \left( \begin{array}{c}
q  \\
\bar{q}_{\varepsilon}
 \end{array} \right)_x+\cdots+b_n\mathcal{R}_{4}^n \left( \begin{array}{c}
q  \\
\bar{q}_{\varepsilon}
 \end{array} \right)_x,
 \end{equation}
where $a_i, b_i$ are some constants, and $i=1,2,\cdots,n$.

As an example we can consider the AKNS system. The AKNS equations \cite{AKNS} can be written as
\begin{equation}\label{AKNSrecursion}
 u_{t}=\mathcal{R} u_x , \,\, \mathrm{where}\,\,  u= \left( \begin{array}{c}
q  \\
r
 \end{array} \right) \,\, \mathrm{and}\,\, \mathcal{R}=\left( \begin{array}{cc}
-qD^{-1}r+\frac{1}{2}D & -qD^{-1}q  \\
rD^{-1}r & rD^{-1}q-\frac{1}{2}D
 \end{array} \right).
 \end{equation}
Here $\mathcal{R}$ is the recursion operator, $D$ is the total $x$-derivative, and $D^{-1}=\int^x$ is the standard anti-derivative. If we consider the nonlocal reduction
\begin{equation}\label{nonreductionAKNS}
r(x,t)=k\bar{q}(\varepsilon_1 x,\varepsilon_2 t)=k\bar{q}_{\varepsilon}, \quad \varepsilon_1^2=\varepsilon_2^2=1,
\end{equation}
where $k$ is a real constant, then the nonlocal reduced system is
\begin{equation}
\left( \begin{array}{c}
q  \\
\bar{q}_{\varepsilon}
 \end{array} \right)_t=\mathcal{R}_{4} \left( \begin{array}{c}
q  \\
\bar{q}_{\varepsilon}
 \end{array} \right)_x
\end{equation}
with the recursion operator
\begin{equation}\label{AKNSnonlocalrecursion}
 \mathcal{R}_{4}=\left( \begin{array}{cc}
-kqD^{-1}\bar{q}_{\varepsilon}+\frac{1}{2}D & -qD^{-1}q  \\
k^2\bar{q}_{\varepsilon}D^{-1}\bar{q}_{\varepsilon} & k\bar{q}_{\varepsilon}D^{-1}q-\frac{1}{2}D
 \end{array} \right).
 \end{equation}
This way we can also generate an infinite hierarchy of nonlocal equations
\begin{equation}
\left( \begin{array}{c}
q  \\
\bar{q}_{\varepsilon}
 \end{array} \right)_{t_{N}}=\mathcal{R}_{4}^N \left( \begin{array}{c}
q  \\
\bar{q}_{\varepsilon}
 \end{array} \right)_x, \quad N=1,2,\cdots,
\end{equation}
and also superposition of integrable nonlocal equations
\begin{equation}
\left( \begin{array}{c}
q  \\
\bar{q}_{\varepsilon}
 \end{array} \right)_t=a_1\mathcal{R}_{4} \left( \begin{array}{c}
q  \\
\bar{q}_{\varepsilon}
 \end{array} \right)_x+a_2\mathcal{R}_{4}^2 \left( \begin{array}{c}
q  \\
\bar{q}_{\varepsilon}
 \end{array} \right)_x+\cdots+a_n\mathcal{R}_{4}^n \left( \begin{array}{c}
q  \\
\bar{q}_{\varepsilon}
 \end{array} \right)_x,
 \end{equation}
where $a_1, a_2,\cdots,a_n$ are some constants and $\mathcal{R}_{4}$ is given by
(\ref{AKNSnonlocalrecursion}). In particular, consider the case for $n=2$,
\begin{equation}\label{superposition}
u_t=\alpha \mathcal{R}u_x+\beta\mathcal{R}^2u_x,\quad u= \left( \begin{array}{c}
q  \\
r
 \end{array} \right),
\end{equation}
where $\alpha$ and $\beta$ are some constants. If we let $\alpha=2a$ and $\beta=-4b$ the above system can be written as
\begin{eqnarray}
&&q_t=aq_{xx}-2aq^2r-bq_{xxx}+6bqrq_x\label{comb1}\\
&&r_t=-ar_{xx}+2aqr^2-br_{xxx}+6bqrr_x.\label{comb2}
\end{eqnarray}
This system of equations has been called as the generalized Hirota equation in \cite{ZZ} which is just the superposition of the coupled nonlinear Schr\"{o}dinger (NLS) and modified Korteweg-de Vries (mKdV) systems. Therefore it should not be considered as a new integrable system of equations.

 If we take $b=0$ then the system becomes the coupled NLS system. We obtained one-, two-, and three-soliton solutions of the coupled NLS system by using the Hirota bilinear method in \cite{GurPek1}. Then we studied all possible local and nonlocal reductions of the coupled NLS system with their soliton solutions. If $a=0$ we have the coupled mKdV system. Similar to the coupled NLS system we gave all the local and nonlocal reductions of the coupled mKdV system with their soliton solutions  in \cite{GurPek2}.

 The local reduction of the system (\ref{comb1}) and (\ref{comb2}) by $r(x,t)=k\bar{q}(x,t)$, $k$ is a real constant, was first introduced by Hirota in \cite{Hirota1973}. He obtained $N$-soliton solutions of this reduced equation  by Hirota bilinear method. In \cite{Demontis} and \cite{Anki} the authors found rogue waves, rational, breather type, and multi-pole solutions of the reduced equation (called the Hirota equation).

 The superposed system (\ref{comb1}) and (\ref{comb2}) has the recursion operator given in (\ref{AKNSrecursion}). When we apply the reduction
(\ref{nonreductionAKNS}) to (\ref{superposition}) with $\alpha=2a$ and $\beta=-4b$ we obtain the nonlocal system
\begin{eqnarray*}
\left( \begin{array}{c}
q  \\
\bar{q}_{\varepsilon}
 \end{array} \right)_t&=&2a\mathcal{R}_{4} \left( \begin{array}{c}
q  \\
\bar{q}_{\varepsilon}
 \end{array} \right)_x-4b\mathcal{R}_{4}^2 \left( \begin{array}{c}
q  \\
\bar{q}_{\varepsilon}
 \end{array} \right)_x\\
 &=&\left( \begin{array}{c}
-2akq^2\bar{q}_{\varepsilon}+aq_{xx}+6bkqq_x\bar{q}_{\varepsilon}-bq_{xxx} \\
2akq(\bar{q}_{\varepsilon})^2-a (\bar{q_{\varepsilon}})_{xx}+6bkq (\bar{q_{\varepsilon}})_x \,\bar{q}_{\varepsilon}-b (\bar{q_{\varepsilon}})_{xxx}.
 \end{array} \right),
\end{eqnarray*}
where $\mathcal{R}_{4}$ is given by (\ref{AKNSnonlocalrecursion}). Clearly, to have a consistent reduction we must have $a=-\bar{a}\varepsilon_2$ and $b=\bar{b}\varepsilon_1\varepsilon_2$.

The nonlocal reduced equation is
\begin{eqnarray}\label{nonlocalsuperposed}
q_t(x,t)&=&aq_{xx}(x,t)-2akq^2(x,t)\bar{q}(\varepsilon_1x,\varepsilon_2t)-bq_{xxx}(x,t)\nonumber\\
&&+6bkq(x,t)q_x(x,t)\bar{q}(\varepsilon_1x,\varepsilon_2t).
\end{eqnarray}
Therefore depending on $(\varepsilon_{1}, \varepsilon_{2})=\{(-1,1),(1,-1),(-1,-1)\}$ we have three type of nonlocal equations:

\noindent \textbf{S-symmetric Hirota equation} $(\varepsilon_1,\varepsilon_2)=(-1,1)$:
\begin{equation}\label{S-symm}
 q_t(x,t)=aq_{xx}(x,t)-akq^2(x,t)\bar{q}(-x,t)-bq_{xxx}(x,t)+6bkq(x,t)q_x(x,t)\bar{q}(-x,t),
\end{equation}
where $a$ and $b$ are pure imaginary numbers.

\noindent \textbf{T-symmetric Hirota equation} $(\varepsilon_1,\varepsilon_2)=(1,-1)$:
\begin{equation}\label{T-symm}
 q_t(x,t)=aq_{xx}(x,t)-akq^2(x,t)\bar{q}(x,-t)-bq_{xxx}(x,t)+6bkq(x,t)q_x(x,t)\bar{q}(x,-t),
\end{equation}
where $a\in \mathbb{R}$ and $b$ is pure imaginary.

\noindent \textbf{ST-symmetric Hirota equation} $(\varepsilon_1,\varepsilon_2)=(-1,-1)$:
\begin{equation}\label{ST-symm}
 q_t(x,t)=aq_{xx}(x,t)-akq^2(x,t)\bar{q}(-x,-t)-bq_{xxx}(x,t)+6bkq(x,t)q_x(x,t)\bar{q}(-x,-t),
\end{equation}
where $a, b\in \mathbb{R}$. All these equations have been proposed by \cite{ZZ} as  new equations.

\section{Soliton solutions}

If we let $ q(x,t)=\frac{g(x,t)}{f(x,t)}$ and $ r(x,t)=\frac{h(x,t)}{f(x,t)}$ in the system (\ref{comb1}) and (\ref{comb2}) we obtain the Hirota bilinear form of the system as
\begin{eqnarray}
&&(D_t-aD_x^2+bD_x^3+\lambda(a-3bD_x))\{g\cdot f\}=0\label{Superh1}\\
&&(D_t+aD_x^2+bD_x^3-\lambda(a+3bD_x))\{h\cdot f\}=0\label{Superh2}\\
&&(D_x^2-\lambda)\{f\cdot f\}=-2gh,\label{Superh3}
\end{eqnarray}
where $\lambda$ is an arbitrary constant. When we apply the Hirota method it arises that $\lambda=0$. Notice that the forms of the one- and two-soliton solutions of the coupled NLS system \cite{GurPek1}, the coupled mKdV system \cite{GurPek2}, and the superposed system (\ref{comb1}) and (\ref{comb2}) are same except the dispersion relations. Indeed we can directly write one-soliton solution of these systems with the pair $(q(x,t),r(x,t))$,
\begin{equation}\label{onesoliton}
\displaystyle q(x,t)=\frac{e^{\theta_1}}{1-\frac{1}{(k_1+k_2)^2} e^{\theta_1+\theta_2}}, \quad \quad r(x,t)=\frac{e^{\theta_2}}{1-\frac{1}{(k_1+k_2)^2}e^{\theta_1+\theta_2}},
\end{equation}
with $ \theta_i=k_ix+\omega_it+\delta_i$, $i=1, 2$, where for the coupled NLS system the dispersion relations are
$\omega_1=ak_1^2, \omega_2=-ak_2^2$; for the coupled mKdV system we have $\omega_i=-bk_i^3, i=1, 2$; and for the superposed system (\ref{comb1}) and (\ref{comb2}) the dispersion relations are
\begin{equation}
\displaystyle \omega_1=ak_1^2-bk_1^3, \quad \omega_2=-ak_2^2-bk_2^3.
\end{equation}
Here $k_{1}$, $k_{2}$, $\delta_{1}$, and $\delta_{2}$ are arbitrary complex numbers.

 Similarly, two-soliton solutions of the coupled NLS \cite{GurPek1}, the coupled mKdV \cite{GurPek2}, and the superposed system (\ref{comb1}) and (\ref{comb2}) are given with the pair $(q(x,t),r(x,t))$,
\begin{eqnarray*}
\displaystyle&& q(x,t)=\frac{e^{\theta_1}+e^{\theta_2}+\gamma_1e^{\theta_1+\theta_2+\eta_1}+\gamma_2e^{\theta_1+\theta_2+\eta_2}}
{1+e^{\theta_1+\eta_1+\alpha_{11}}+e^{\theta_1+\eta_2+\alpha_{12}}+e^{\theta_2+\eta_1+\alpha_{21}}+e^{\theta_2+\eta_2+\alpha_{22}}
+Me^{\theta_1+\theta_2+\eta_1+\eta_2}},\nonumber\\
&&r(x,t)=\frac{e^{\eta_1}+e^{\eta_2}+\beta_1e^{\theta_1+\eta_1+\eta_2}+\beta_2e^{\theta_2+\eta_1+\eta_2}}
{1+e^{\theta_1+\eta_1+\alpha_{11}}+e^{\theta_1+\eta_2+\alpha_{12}}+e^{\theta_2+\eta_1+\alpha_{21}}+e^{\theta_2+\eta_2+\alpha_{22}}
+Me^{\theta_1+\theta_2+\eta_1+\eta_2}},\\
&\label{twosoliton}
\end{eqnarray*}
where
\begin{eqnarray*}
&&\displaystyle e^{\alpha_{ij}}=-\frac{1}{(k_i+\ell_j)^2}, \quad \gamma_i=-\frac{(k_1-k_2)^2}{(k_1+\ell_i)^2(k_2+\ell_i)^2},\\ &&\beta_i=-\frac{(\ell_1-\ell_2)^2}{(\ell_1+k_i)^2(\ell_2+k_i)^2},\quad
M=\frac{(k_1-k_2)^2(l_1-l_2)^2}{(k_1+l_1)^2(k_1+l_2)^2(k_2+l_1)^2(k_2+l_2)^2},
\end{eqnarray*}
and $\displaystyle \theta_i=k_ix+\omega_it+\delta_i$, $\displaystyle \eta_i=\ell_ix+m_it+\alpha_i$ for  $1\leq i,j\leq 2$, where
for the coupled NLS system the dispersion relations are $\omega_i=ak_i^2, m_i=-a\ell_i^2, i=1, 2$; for the coupled mKdV system we have $\omega_i=-bk_i^3, m_i=-b\ell_i^3, i=1, 2$; and for the superposed system (\ref{comb1}) and (\ref{comb2}) the dispersion relations are
\begin{equation}\displaystyle
\omega_i=ak_i^2-bk_i^3,\quad m_i=-a\ell_i^2-b\ell_i^3\quad i=1, 2.
\end{equation}
Here $k_{i}$, $\ell_{i}, \delta_{i}$, and $\alpha_{i}$, $i=1, 2$ are arbitrary complex numbers.

In \cite{ZZ}, the authors also obtained the one- and two-soliton solutions of the superposed system of the coupled NLS and mKdV systems (\ref{comb1}) and (\ref{comb2}) by using the Hirota method but these solutions have already been given  previously in \cite{GurPek1}-\cite{GurPek3}  except the dispersion relations.
In addition, the authors studied one- and two-soliton solutions of the nonlocal reductions of the superposed system (\ref{comb1}) and (\ref{comb2}) in \cite{ZZ}. Obviously, one also does not need to find the soliton solutions of the nonlocal reductions of the superposed system (\ref{comb1}) and (\ref{comb2}) as well since these nonlocal reductions are also superpositions of the nonlocal reductions of the coupled NLS and mKdV systems which have already studied in \cite{GurPek1} and \cite{GurPek2}. We derived one- and two-soliton solutions of the nonlocal reduced equations obtained by the reduction $r(x,t)=\bar{q}(\varepsilon_1x,\varepsilon_2t)$ in \cite{GurPek1} and \cite{GurPek2}, as
\begin{equation}\label{onesolitonnonlocal}
q(x,t)=\frac{e^{k_1x+\omega_1t+\delta_1}}{1-\frac{k}{(k_1+\bar{k}_1\varepsilon_1)^2}e^{(k_1+\bar{k}_1\varepsilon_1)x+(\omega_1+\omega_2)t+\delta_1+\bar{\delta}_1}    },
\end{equation}
where for nonlocal NLS equations the dispersion relations are $\omega_1=ak_1^2, \omega_2=-a\bar{k}_1^2$ and for nonlocal mKdV equations we have $
\omega_1=-bk_1^3, \omega_2=-b\varepsilon_1\bar{k}_1^3$, with $a=-\bar{a}\varepsilon_2, b=\bar{b}\varepsilon_1\varepsilon_2$. Therefore one-soliton solution of the nonlocal superposed equations (\ref{nonlocalsuperposed}) is again (\ref{onesolitonnonlocal}) with the dispersion relations
\begin{equation}
\omega_1=ak_1^2-bk_1^3,\quad \omega_2=-a\bar{k}_1^2-b\varepsilon_1\bar{k}_1^3,\quad \mathrm{for}\quad a=-\bar{a}\varepsilon_2, \quad b=\bar{b}\varepsilon_1\varepsilon_2.
\end{equation}

\section{Conclusion}
The aim of this  work is to remark on two basic wrong approaches. The first one is  proposing the superpositions of integrable hierarchy of equations as new integrable systems and the second one is obtaining the soliton solutions of these superposed equations by the  Hirota bilinear method.  For this purpose we gave the superposition of the coupled NLS and mKdV systems, as an example,  which are members of AKNS hierarchy with well-known recursion operator.
These systems are integrable and  the superposition of them is also integrable possessing the same recursion operator. We pointed out that the
superposed systems should not be considered as a new integrable system, because they are integrable by construction.
The soliton solutions of the coupled NLS and mKdV and the superposed systems obtained by Hirota bilinear
method are exactly same except their dispersion relations. We have the same conclusion for the  local and nonlocal reductions of the coupled NLS and mKdV systems. Their  superposition does not give any new integrable equations and furthermore the reduced superposed equations have the same form of soliton solutions with different  dispersion relations.


\end{document}